\documentstyle[12pt]{article}
\newcommand{\be}{\begin{equation}}
\newcommand{\ee}{\end{equation}}
\newcommand{\bit}{\begin{itemize}}
\newcommand{\eit}{\end{itemize}}
\newcommand{\bd}{\begin{displaymath}}
\newcommand{\ed}{\end{displaymath}}
\newcommand{\baa}{\begin{array}{lll}}
\newcommand{\eaa}{\end{array}}
\newcommand{\ba}{\begin{eqnarray}}
\newcommand{\ea}{\end{eqnarray}}
\newcommand{\la}{\label}

\def\GeV{{\rm GeV}}

\def \as{\relax\ifmmode\alpha_s\else{$\alpha_s${ }}\fi}

\begin{document}
\begin{center}
{\large\bf THE PHOTON STRUCTURE FUNCTION $F_2$ IN QCD WITH
 NONLOCAL VACUUM  QUARK CONDENSATES }
\end{center}
\begin{center}
A.P. BAKULEV
\footnote{VINITI RAN, Moscow, Russia;
E-mail bakulev@thsun1.jinr.dubna.su}
and
S.V. MIKHAILOV
\footnote{E-mail mikhs@thsun1.jinr.dubna.su;
on leave of absence from Rostov State University, Rostov-Don, Russia}\\
{\em Bogoliubov Laboratory of Theoretical Physics, JINR, Dubna, Russia}
\end{center}
\begin{abstract}
{\footnotesize
We calculate the contribution from nonlocal vacuum condensates of quark
fields to the hadronic part of the photon structure function $F_2(x)$ in
the operator product expansion approach to QCD and as a result obtain a
substantial improvement of the agreement with experimental data for the
standard value of the parameter
$\lambda_q^2 \equiv \langle\bar qD^2 q\rangle /\langle\bar qq\rangle
\approx 0.5 \ \ GeV^2 $.}
\end{abstract}

1. Here we present  the calculation of the hadronic part of the photon
structure function (PhSF) by taking into account the nonlocal quark
condensate and following the method suggested by Gorsky et al. \cite{GI90}.
We also compare our results with  experiments and provide a brief
discussion.

  The approach to calculate the hadronic part of the PhSF $F_2(x,P^2,Q^2)$
in photon - photon DIS
($Q^2 \gg P^2$) in the region of moderate values of $x$ and  $Q^2$ has been
proposed in \cite{GI90}. This approach is based on the operator product
expansion (OPE) technique for a 4-current correlator with account of the
non-zero vacuum condensate (VC) of gluon fields $(\langle GG\rangle)$.
The authors of this paper have managed to reproduce rather well the existing
experimental data in the region $0.3 < x < 0.7$ for different values of $Q^2$
($Q^2 = 4.3,  5.3, 9.2, 23.0\, \GeV^2$) {\bf without introducing new
phenomenological parameters}. As to VC of quark fields, it was asserted that
their contribution (through the operators of the lowest dimension, i.e.
$:\bar qq:$) is proportional to the $\delta(1-x)$ and thus, could be dropped
out of the consideration of the region $0.3<x<0.7$.
(The contributions of quark VCs to PhSF in this region are possible also from
diagrams with radiative corrections to the usual box diagrams, which are of
order $\alpha_S$ and for this reason aren't considered.)

     However, the use of the nonlocal quark condensate \cite{MR89}, which is
equivalent to the summation of an infinite set of condensates of higher
dimensions, leads to the result with new properties. We show that nonlocal
quark VCs give rise to a smooth (over $x$) contribution to PhSF. It is
interesting to note, that essential diagrams are new. The magnitude of the
corresponding contribution to PhSF ($\Delta^{\langle\bar qq\rangle}F_2$) is
of the same order as the one from gluon VC ($\Delta^{\langle GG\rangle}F_2$)
and is determined by the characteristic length of the quark vacuum
correlations  $1/\lambda_q$:
\ba \la{VC}
\Delta^{\langle GG\rangle}F_2(x) \sim
\frac{1}{m_{\rho}^4}\langle\frac{\alpha_S}{\pi}GG\rangle,
{}~~~~ \Delta^{\langle\bar qq\rangle} F_2(x) \sim
\frac{\pi^4\langle\bar qq\rangle^2}{\lambda_q^6}.
\ea
The latter estimate in (\ref{VC}) doesn't depend on the concrete form of a
condensate distribution function \cite{MR89} and reflects the essentially
nonperturbative character of this contribution. Numerical values are:
$\Delta^{\langle\bar qq\rangle}F_2 \approx 0.35$, whereas
$\Delta^{\langle GG\rangle}F_2 \approx -0.15$  in the central region of $x$
for the standard values $\lambda_q^2 \approx 0.4 \, \GeV^2 \
(\lambda_q^2 \equiv \langle\bar qD^2 q\rangle /\langle\bar qq\rangle
\approx 0.4 - 0.6 \ \GeV^2)$ \cite{belioffe}.
Careful analysis shows the substantial improvement of agreement with
experiment in all the region $0.2 < x < 0.8$ just for these standard values
of correlation length $1/\lambda_q$. For larger values,
$\lambda_q^2 \approx 1.2 \, \mbox{GeV}^2$, which are typical of instanton
liquid models \cite{shuryak89}, the quark VC's contribution is of no
importance as compared with the gluon one. It's clear however, that the value
of
$\lambda_q$ couldn't be much less than the standard value because of the
"explosion" of $\Delta^{\langle\bar qq\rangle}F_2$ for $\lambda_q \rightarrow
0$.
One can conclude that the PhSF is rather sensitive to the parameters of
condensate structure, therefore the extraction of $F_2$ with high precision
from experiments on photon-photon interactions provides the possibility of
independent evaluation of the length of correlations in the nonperturbative
QCD vacuum.

     We also establish the breakdown of factorization theorem \cite{echaja1}
for the discontinuities (i.e. imaginary parts, {\em Disc}) of diagrams with
nonlocal VC on the cut line. This effect depends on the decay rate of vacuum
correlations at large distances: {\it e.g.}, the Gauss decay of quark VC
($\langle\bar q(z)q(0)\rangle \sim \exp(-\gamma z^2)$ for $ |z| \rightarrow
\infty $) does violate the factorization, but the exponential one
($\sim \exp(-\gamma |z|)$)  doesn't. At the same time, our conclusion about
the necessity to take into account $\delta$\--function distributions isn't
related with the nonlocality of VCs: the distribution concentrated near the
border of some area couldn't be used properly for treating local problems in
the center of the area. In this case the appearance of a border-concentrated
distribution like the $\delta$--function from the OPE signals about the
deficiency of that expansion. Our method, that uses another type of
distributions, avoids this problem.

2. The nonlocal VC seems to be introduced for the first time in \cite{nlc},
and the exponential decay in coordinate representation was obtained in
lattice calculations \cite{lat84}. The nonlocal VC was successfully employed
to explain different dynamical hadron properties in exclusive processes
\cite{MR90BR91}. We use a $\delta$\--shaped Ansatz for the distribution
functions $f(\alpha)$ of nonlocal scalar ($M(z)$) and vector ($M_\mu (z)$)
quark VCs \cite{MR92}:
\ba
\label{eq:qSq}
M(z)\equiv \langle\bar q(0)\hat E(0,z)q(z)\rangle =
\langle\bar q(0)q(0)\rangle \int_{0}^{\infty} e^{\alpha z^2/4}\,
f_S(\alpha)\, d\alpha ; \\
\label{eq:qVq}
M_\mu (z)\equiv\langle\bar q(0)\gamma_\mu \tilde E(0,z)q(z)\rangle =
-iz_\mu A\int_{0}^{\infty} e^{\alpha z^2/4}\, f_V(\alpha)\, d\alpha ; \\
\label{eq:fmod}
f_S^{mod}(\alpha) = \delta\left(\alpha-\Delta^2\right); \qquad
f_V^{mod}(\alpha) = \delta_{\alpha}^{'} \left(\alpha-\Delta_V^2\right) ,
\ea
where $A = 2/81 \pi \alpha_s \langle\bar qq\rangle^2$, and scales
$\Delta^2 \equiv \lambda_q^2/2$ and
$\Delta_V^2 \equiv a_V \lambda_q^2/2$
aren't always equal, \ $ \langle \sqrt{\alpha_S}\bar qq\rangle^{1/3}
\approx 0.23 $ \ GeV.
 We select the value $a_V = 0.7$ , that is consistent
with the Taylor expansion of the VC in the lowest orders \cite{MR89}.

  Let us consider diagrams of fig.1 (crossed graphs are also understood;
the contribution of diagram 1A is marked by index $S$, and that of 1B --
by $V$) and take into account the contributions of the lowest twist,
i.e. neglect terms  like
$ P^2 / Q^2,\ \ \exp\left(-Q^2/(\Delta^2x)\right)$. Then, denoting
$$C_{norm} \equiv \frac{3\alpha_{em}\sum_{q}e_q^4}{\pi};\, \,
\gamma_S \equiv \frac{2xP^2}{\Delta^2};\, \,
\gamma_V \equiv \frac{\bar xP^2}{\Delta_V^2},$$
we obtain
\ba
\frac{1}{C_{norm}}\Delta_S^{\langle\bar qq\rangle}F_2^T\left(x;P^2\right) &=&
\frac{8\pi^4}{9\Delta^6} \langle\bar qq\rangle^2
\left(x\bar x \right) \exp\left(-\gamma_S\right);
\label{scalarT} \\
\frac{1}{C_{norm}}\Delta_S^{\langle\bar qq\rangle}F_2^L\left(x;P^2\right) &=&
\frac{8\pi^4}{9\Delta^6} \langle\bar qq\rangle^2
\left(\frac{x^2}{\gamma_S}\right) \exp\left(-\gamma_S\right);
\label{scalarL} \\
 \frac{1}{C_{norm}}\Delta_V^{\langle\bar qq\rangle}F_2^T\left(x;P^2\right) &=
 &-\frac{32\pi^3}{81\Delta_V^6} \alpha_S\langle\bar qq\rangle^2
x \exp\left(-\gamma_V\right) \times \nonumber \\
&& \times \left(x^2+\bar x^2 -\gamma_V\left(x^2+\frac{x+1}{2} \right)
+\gamma_V^2\frac{x}{2} \right);
\label{vectorT} \\
 \frac{1}{C_{norm}}\Delta_V^{\langle\bar qq\rangle}F_2^L\left(x;P^2\right) &=&
-\frac{32\pi^3}{81\Delta_V^6} \alpha_S\langle\bar qq\rangle^2
x\bar x \exp\left(-\gamma_V\right)  \times  \nonumber \\
&& \times\left(\frac{2x-1}{2\gamma_V}-3x-\frac{1}{2}+\gamma_V\frac{5x+1}{2}
-\gamma_V^2\frac{x}{2} \right).
\label{vectorL}
\ea
\begin{figure}[h]
\unitlength=2.25pt
\special{em:linewidth 0.75pt}
\linethickness{0.75pt}
\begin{picture}(189.00,52.00)
\put(28.00,8.67){\oval(2.00,2.00)[lt]}
\put(28.00,10.17){\oval(2.00,1.00)[rb]}
\put(30.00,10.67){\oval(2.00,2.00)[lt]}
\put(30.00,12.17){\oval(2.00,1.00)[rb]}
\put(32.00,12.67){\oval(2.00,2.00)[lt]}
\put(32.00,14.17){\oval(2.00,1.00)[rb]}
\put(68.00,10.67){\oval(2.00,2.00)[lb]}
\put(68.00,8.67){\oval(2.00,2.00)[rt]}
\put(66.00,12.67){\oval(2.00,2.00)[lb]}
\put(66.00,10.67){\oval(2.00,2.00)[rt]}
\put(64.00,14.67){\oval(2.00,2.00)[lb]}
\put(64.00,12.67){\oval(2.00,2.00)[rt]}
\put(68.00,10.67){\oval(2.00,2.00)[lb]}
\put(68.00,8.67){\oval(2.00,2.00)[rt]}
\put(66.00,12.67){\oval(2.00,2.00)[lb]}
\put(66.00,10.67){\oval(2.00,2.00)[rt]}
\put(64.00,14.67){\oval(2.00,2.00)[lb]}
\put(64.00,12.67){\oval(2.00,2.00)[rt]}
\put(68.00,12.67){\makebox(0,0)[cc]{$p$}}
\put(48.00,3.00){\makebox(0,0)[cb]{A}}
\put(63.00,30.00){\makebox(0,0)[cc]{$f_S$}}
\put(68.00,46.00){\makebox(0,0)[cc]{$q$}}
\put(33.00,15.00){\line(1,0){30.11}}
\put(33.00,45.00){\line(1,0){30.11}}
\put(63.11,45.11){\line(0,-1){10.00}}
\put(63.00,15.00){\line(0,1){0.11}}
\put(63.00,25.00){\circle*{4.00}}
\put(63.00,35.00){\circle*{4.00}}
\put(63.00,30.00){\oval(12.00,20.00)[]}
\put(33.00,30.00){\makebox(0,0)[cc]{$f_S$}}
\put(28.00,13.00){\makebox(0,0)[ct]{$p$}}
\put(33.11,45.11){\line(0,-1){10.00}}
\put(33.00,15.00){\line(0,1){0.11}}
\put(33.00,25.00){\circle*{4.00}}
\put(33.00,35.00){\circle*{4.00}}
\put(33.00,30.00){\oval(12.00,20.00)[]}
\put(64.00,44.67){\oval(2.00,2.00)[lt]}
\put(64.00,46.17){\oval(2.00,1.00)[rb]}
\put(66.00,46.67){\oval(2.00,2.00)[lt]}
\put(66.00,48.17){\oval(2.00,1.00)[rb]}
\put(68.00,48.67){\oval(2.00,2.00)[lt]}
\put(68.00,50.17){\oval(2.00,1.00)[rb]}
\put(32.00,46.67){\oval(2.00,2.00)[lb]}
\put(32.00,44.67){\oval(2.00,2.00)[rt]}
\put(30.00,48.67){\oval(2.00,2.00)[lb]}
\put(30.00,46.67){\oval(2.00,2.00)[rt]}
\put(28.00,50.67){\oval(2.00,2.00)[lb]}
\put(28.00,48.67){\oval(2.00,2.00)[rt]}
\put(28.00,46.00){\makebox(0,0)[cc]{$q$}}
\put(33.00,15.00){\line(0,1){10.11}}
\put(63.00,25.00){\line(0,-1){9.89}}
\put(88.00,8.67){\oval(2.00,2.00)[lt]}
\put(88.00,10.17){\oval(2.00,1.00)[rb]}
\put(90.00,10.67){\oval(2.00,2.00)[lt]}
\put(90.00,12.17){\oval(2.00,1.00)[rb]}
\put(92.00,12.67){\oval(2.00,2.00)[lt]}
\put(92.00,14.17){\oval(2.00,1.00)[rb]}
\put(128.00,10.67){\oval(2.00,2.00)[lb]}
\put(128.00,8.67){\oval(2.00,2.00)[rt]}
\put(126.00,12.67){\oval(2.00,2.00)[lb]}
\put(126.00,10.67){\oval(2.00,2.00)[rt]}
\put(124.00,14.67){\oval(2.00,2.00)[lb]}
\put(124.00,12.67){\oval(2.00,2.00)[rt]}
\put(128.00,10.67){\oval(2.00,2.00)[lb]}
\put(128.00,8.67){\oval(2.00,2.00)[rt]}
\put(126.00,12.67){\oval(2.00,2.00)[lb]}
\put(126.00,10.67){\oval(2.00,2.00)[rt]}
\put(124.00,14.67){\oval(2.00,2.00)[lb]}
\put(124.00,12.67){\oval(2.00,2.00)[rt]}
\put(128.00,12.67){\makebox(0,0)[cc]{$p$}}
\put(108.00,3.00){\makebox(0,0)[cb]{B}}
\put(128.00,46.00){\makebox(0,0)[cc]{$q$}}
\put(93.00,45.00){\line(1,0){30.11}}
\put(123.00,15.00){\line(0,1){0.11}}
\put(88.00,13.00){\makebox(0,0)[ct]{$p$}}
\put(93.00,15.00){\line(0,1){0.11}}
\put(124.00,44.67){\oval(2.00,2.00)[lt]}
\put(124.00,46.17){\oval(2.00,1.00)[rb]}
\put(126.00,46.67){\oval(2.00,2.00)[lt]}
\put(126.00,48.17){\oval(2.00,1.00)[rb]}
\put(128.00,48.67){\oval(2.00,2.00)[lt]}
\put(128.00,50.17){\oval(2.00,1.00)[rb]}
\put(92.00,46.67){\oval(2.00,2.00)[lb]}
\put(92.00,44.67){\oval(2.00,2.00)[rt]}
\put(90.00,48.67){\oval(2.00,2.00)[lb]}
\put(90.00,46.67){\oval(2.00,2.00)[rt]}
\put(88.00,50.67){\oval(2.00,2.00)[lb]}
\put(88.00,48.67){\oval(2.00,2.00)[rt]}
\put(88.00,46.00){\makebox(0,0)[cc]{$q$}}
\put(108.00,15.00){\makebox(0,0)[cc]{$f_V$}}
\put(93.00,45.00){\line(0,-1){0.11}}
\put(93.11,15.11){\line(1,0){10.00}}
\put(123.00,45.00){\line(0,-1){0.11}}
\put(123.11,15.11){\line(-1,0){10.00}}
\put(103.00,15.00){\circle*{4.00}}
\put(113.00,15.00){\circle*{4.00}}
\put(108.00,15.00){\oval(20.00,12.00)[]}
\put(148.00,8.67){\oval(2.00,2.00)[lt]}
\put(148.00,10.17){\oval(2.00,1.00)[rb]}
\put(150.00,10.67){\oval(2.00,2.00)[lt]}
\put(150.00,12.17){\oval(2.00,1.00)[rb]}
\put(152.00,12.67){\oval(2.00,2.00)[lt]}
\put(152.00,14.17){\oval(2.00,1.00)[rb]}
\put(188.00,10.67){\oval(2.00,2.00)[lb]}
\put(188.00,8.67){\oval(2.00,2.00)[rt]}
\put(186.00,12.67){\oval(2.00,2.00)[lb]}
\put(186.00,10.67){\oval(2.00,2.00)[rt]}
\put(184.00,14.67){\oval(2.00,2.00)[lb]}
\put(184.00,12.67){\oval(2.00,2.00)[rt]}
\put(188.00,10.67){\oval(2.00,2.00)[lb]}
\put(188.00,8.67){\oval(2.00,2.00)[rt]}
\put(186.00,12.67){\oval(2.00,2.00)[lb]}
\put(186.00,10.67){\oval(2.00,2.00)[rt]}
\put(184.00,14.67){\oval(2.00,2.00)[lb]}
\put(184.00,12.67){\oval(2.00,2.00)[rt]}
\put(188.00,12.67){\makebox(0,0)[cc]{$p$}}
\put(168.00,3.00){\makebox(0,0)[cb]{B$_{\rm fact}$}}
\put(188.00,46.00){\makebox(0,0)[cc]{$q$}}
\put(153.00,45.00){\line(1,0){30.11}}
\put(183.00,15.00){\line(0,1){0.11}}
\put(148.00,13.00){\makebox(0,0)[ct]{$p$}}
\put(153.00,15.00){\line(0,1){0.11}}
\put(184.00,44.67){\oval(2.00,2.00)[lt]}
\put(184.00,46.17){\oval(2.00,1.00)[rb]}
\put(186.00,46.67){\oval(2.00,2.00)[lt]}
\put(186.00,48.17){\oval(2.00,1.00)[rb]}
\put(188.00,48.67){\oval(2.00,2.00)[lt]}
\put(188.00,50.17){\oval(2.00,1.00)[rb]}
\put(152.00,46.67){\oval(2.00,2.00)[lb]}
\put(152.00,44.67){\oval(2.00,2.00)[rt]}
\put(150.00,48.67){\oval(2.00,2.00)[lb]}
\put(150.00,46.67){\oval(2.00,2.00)[rt]}
\put(148.00,50.67){\oval(2.00,2.00)[lb]}
\put(148.00,48.67){\oval(2.00,2.00)[rt]}
\put(148.00,46.00){\makebox(0,0)[cc]{$q$}}
\put(168.00,15.00){\makebox(0,0)[cc]{$f_V$}}
\put(153.00,45.00){\line(0,-1){0.11}}
\put(153.11,15.11){\line(1,0){10.00}}
\put(183.00,45.00){\line(0,-1){0.11}}
\put(183.11,15.11){\line(-1,0){10.00}}
\put(163.00,15.00){\circle*{4.00}}
\put(173.00,15.00){\circle*{4.00}}
\put(168.00,15.00){\oval(20.00,12.00)[]}
\put(93.00,45.00){\line(0,-1){29.89}}
\put(123.00,45.00){\line(0,-1){29.89}}
\put(153.00,45.00){\line(-3,-5){2.78}}
\put(186.00,40.00){\line(-3,5){2.89}}
\put(153.00,15.00){\line(2,3){14.78}}
\put(167.78,37.11){\line(2,-3){14.67}}
\put(168.00,37.00){\makebox(0,0)[cc]{$\otimes$}}
\put(132.00,30.00){\vector(1,0){12.00}}
\put(48.00,8.00){\line(0,1){4.00}}
\put(48.00,14.00){\line(0,1){4.00}}
\put(48.00,20.00){\line(0,1){4.00}}
\put(48.00,26.00){\line(0,1){4.00}}
\put(48.00,32.00){\line(0,1){4.00}}
\put(48.00,38.00){\line(0,1){4.00}}
\put(48.00,44.00){\line(0,1){4.00}}
\put(48.00,50.00){\line(0,1){2.00}}
\put(108.00,8.00){\line(0,1){4.00}}
\put(108.00,14.00){\line(0,1){4.00}}
\put(108.00,20.00){\line(0,1){4.00}}
\put(108.00,26.00){\line(0,1){4.00}}
\put(108.00,32.00){\line(0,1){4.00}}
\put(108.00,38.00){\line(0,1){4.00}}
\put(108.00,44.00){\line(0,1){4.00}}
\put(108.00,50.00){\line(0,1){2.00}}
\put(168.00,40.00){\line(0,1){3.00}}
\put(168.00,44.00){\line(0,1){3.00}}
\put(168.00,48.00){\line(0,1){3.00}}
\end{picture}
\caption{ Diagrams with nonlocal VCs, essential for PhSF $F_2(x)$.
To the left there is the diagram with scalar VCs; in the middle, the
initial diagram with
vector VC, to the right, its factorized form.}
\label{fig:box-nlc}
\end{figure}
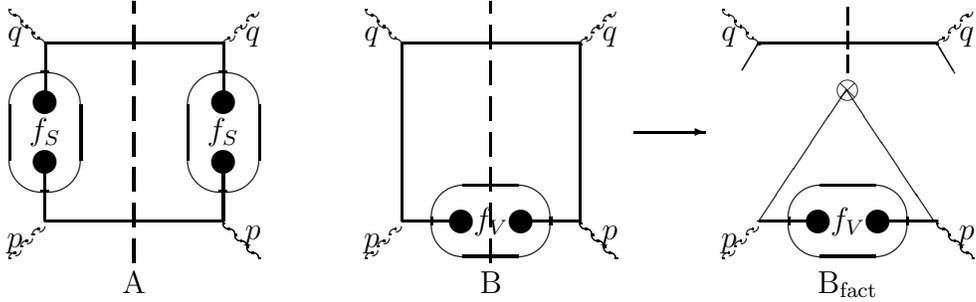

  It should be noted that the methods used for calculating these
contributions are different: for the scalar VCs the Cutkosky's approach was
applied (it is equivalent to the double Borel transform method, for details
see \cite{AB91}), and for the vector VC we used the  factorization of
large and small distance contributions directly -- the {\em Disc} of diagram
 is determined the by coefficient function of the process
(i.e., at short distances). This difference is explained by the
following. The exact calculation of the imaginary part of diagram 1B with the
nonlocal vector VC with the $\delta$ - Ansatz form (\ref{eq:qVq}) gives
the exact zero. This is a simple consequence of the more general

\vspace{0.2cm}

\parbox{6.4in}{{\bf Proposition}:

   {\em If the weight $f(\alpha)$ of some line of a diagram in $\alpha$-space
   is concentrated on the bounded support,
   then the contribution to Disc of this diagram from the cut through this
   line is zero.}}\footnote {We are grateful to O. Teryaev for stimulating
   discussion on this point.}
\vspace{0.4cm}

\noindent This can be proved by direct calculations in the case of ``box"
diagrams. The distribution $f_V^{mod}(\alpha)$ (\ref{eq:fmod})  we use
satisfies
conditions of this proposition. Meanwhile there is a whole class of Ansatzes
$h_{n}(\alpha)$, for which the nonlocal VCs:
\bit
\item {decay in the large $|z|$ limit exponentially
($\sim \exp(-\gamma |z|)$);}
\item {have the unbounded support in $\alpha$\--space, e.g. :
\ba
h_{n}(\alpha) = exp(\frac{-m_n^2}{\alpha})\cdot
\frac{\alpha^{-n}(m_n^2)^{n-1}}
{\Gamma(n-1)} ;
\nonumber
\ea }
\item {imitate the $\delta$\--shaped Ansatz for large values of the
 parameter $n$;}
\eit
The imaginary part of diagram 1B for these Ansatzes isn't zero. But
this quantity is essentially Ansatz-dependent. For this reason we use
the factorization method, which is weak sensitive to the concrete choice
of Ansatz (we could use the Ansatz $h_{n}(\alpha)$ in all the calculations,
but it would produce substantially complicated expressions and numerically
not rather different results).

We see, that the nonlocal quark VCs gives smooth  (over $x$)
contributions in all the region $0<x<1$ and the parameter of nonlocality is
 located in the denominators of a common factors, which is a  signal of
nonperturbativness of these corrections. We also want to emphasize that
in the limit $P^2 \rightarrow 0$ only contributions to $F_2^L$ are singular
whereas those related to $F_2^T$ are regular.

3. For treating SF of a real photon it's necessary to realize in some way
an extrapolation of the result
obtained for $P^2 \gg \Lambda_{QCD}^2$
to the region $P^2 \rightarrow 0$.
First of all, we should remember that in this limit physical
SF $F_2^T$ and $F_2$  coincide, and physical $F_2^L \rightarrow 0$ (for
details see \cite{GI90}).
So, we'll consider further only the transverse part of PhSF, i.e. $F_2^T$.
Then, we can extrapolate to $P^2 = 0$ our results for nonlocal quark VC
contributions without any problem.
Note that we don't pretend to describe the region of large $P^2 \gg \Delta^2$
in (\ref{scalarT})-(\ref{vectorL}), because this asymptotic regime is
determined by the unknown details of distribution function $f(\alpha)$
in large $\alpha$ region.
For treating singularities like $1/P^4$ (from the gluon VC) and
$\log( Q^2 / P^2)$ (from the perturbative part) which appear in OPE
calculations in QCD,  we used the method and model of \cite{GI90}.
By this method, the PhSF is represented via dispersion relations in $p^2$
in terms of the contributions of physical states (vector meson ($\rho$) +
 continuum) and
the parameters of the model are chosen so that they
correctly reproduce all the terms of OPE calculations.

 Then for the real PhSF we obtain:
\ba
\la{final}
\frac{1}{C_{norm}}F_2(x)  =  x\left\{-1+6x\bar x +
\left[x^2+\bar x^2\right] \log\left(\frac{Q^2}{x^2p_0^2}\right)
 + \right.
\nonumber \\
 \left.\left[  \frac{8\pi^4}{9\Delta^6} \langle\bar qq\rangle^2 \bar x
 - \left[x^2+\bar x^2\right] \frac{32\pi^3}{81\Delta^6}
  \alpha_S\langle\bar qq\rangle^2 \right]
  - \
\frac{p_0^4}{2m_{\rho}^4} \left[x^2+\bar x^2+
       \frac{8\pi^2}{27p_0^4x^2}\langle\frac{\alpha_S}{\pi}GG\rangle \right]
 \   \right\}
\ea
where $p_0^2 \approx 1.5 \GeV$ is the standard value of continuum threshold
in the QCD sum rule calculations of the $\rho$-meson properties and $m_{\rho}$
is the $\rho$-meson mass. The quark contribution to hadronic part is shown in
the first term of the second line in (\ref{final}).
To compare the new result (\ref{final}) with experimental data \cite{PLUTO},
\cite{TASSO}, we should include
the evolution of the quark VC with $Q^2$ :
$ \langle\bar qq\rangle^2 (\mu^2 \sim  \GeV^2) \to
\langle\bar qq\rangle^2 (Q^2), \langle\bar q g (\sigma G) q\rangle(\mu^2) \to
\langle\bar q g (\sigma G) q\rangle(Q^2)$
following usual one loop evolution formulae (e.g. \cite{MOR84}) .
The quark terms lead to the growth of the hadronic part in the central
region of $x$ and happily works to the better agreement with the experiment,
as represented in Fig. 2. Moreover, in the region of $x \geq 0.8 $
there is a weak tendency, due to quark VC corrections, to the lower growth of
the curve. The best agreement with data is achieved for the value of
$\lambda_q^2 = 0.5$ -- $0.6 \ \GeV^2$.

Nevertheless, the experimental errors should be reduced few times for
comparing experiment with theory curve carefully.
Further theoretical progress may be reached in the three ways:
\begin{enumerate}
\item To improve the hadronic model of PhSF (see  \cite{GI90}) by
 introducing  new vector resonances.
 This step can improve the behavior of the theoretical curve at ``large''
 $x \geq 0.7$ .
\item To revise the expression for gluon corrections by taking into account
 the correlation length
  of gluon condensate \cite{SD88S89}, \cite{MS93}. It will correct the
  theoretical curve in the region  $x \sim 0.2$.
\item At $x \sim 0$, i.e. near the singularity in $t$--channel,
     the OPE series diverges \cite{GI90}. But it may be corrected by
     reformation
     \footnote{Authors are grateful to R.Ruskov for stressing this point.}
      of OPE in the way of cancellation of the ``long distance''
     contribution following the approach in \cite{RR93}

  \end{enumerate}

The authors would like to thank Prof. A.V. Radyushkin for pointing the
problem, stimulating discussions and kind support, and also Prof. B.L.
Ioffe,  Drs. V.M. Braun, R. Ruskov and O.V. Teryaev for valuable discussions.

This work was supported in part by the Russian Foundation for
Fundamental Researches Grant $N^0$ 93-02-3811
and in part by the Grant $N^0$ RFE 000 from International Science Foundation.


\newpage
\begin{center}\underline {Figure 2:} \end{center}
\noindent Comparison of the theoretical model predictions with  experimental
data for $Q^2 = 4.3$ GeV$^2,$ $Q^2 = 5.3$ GeV$^2$ and
$Q^2 = 9.2$ GeV$^2$ from \protect{\cite{PLUTO}} and for
$Q^2 = 23.0$ GeV$^2$ from \protect{\cite{TASSO}}.
 Solid line - our results; dashed - results of \protect{\cite{GI90} and
 the lowest dashed (long dashes) -- hadronic part contribution.
\end{document}